\documentclass[12pt]{spieman} % 12pt font required by SPIE;
\usepackage{amsmath,amsfonts,amssymb}
\usepackage{graphicx}
\usepackage{setspace}
\usepackage{tocloft}
\usepackage{lineno}
\usepackage{multirow}
\usepackage{hyperref}
\title{An Analysis of Large Astronomical Detector Controller Systems and Implications for Future ESO Detector Systems}

\author[a]{Mathias Richerzhagen}
\author[a]{Naidu Bezawada}
\author[a]{Sebastian Elias Egner}
\author[a]{Elizabeth George}
\author[a]{Alessandro Meoli}
\author[a]{Alexander Rüde}
\author[a]{Matthias Seidel}
\author[a]{Domingo \'Alvarez M\'endez}
\author[a]{Olaf Iwert}
\author[a]{Leander Mehrgan}
\author[a]{Javier Reyes}
\author[a]{Beno\^it Serra}
\author[a]{Omar Sqalli}
\author[a]{Derek Ives}
\affil[a]{European Southern Observatory (ESO), Karl-Schwarzschild-Str. 2, 85748 Garching bei München, Germany}

\cftpagenumbersoff{figure}
\cftpagenumbersoff{table} 
\begin{document} 
\maketitle

\begin{abstract}
Large astronomical instruments using tens to hundreds of optical or infrared science detectors pose specific challenges for detector control, where{,} in addition to performance, other engineering aspects like scalability, power consumption, size{,} weight and programmatic aspects such as cost and sustainability need to be considered. In this paper we analyze the approach existing instruments have taken for detector control. We focus this analysis on recent ground based astronomical instruments using 10 or more detectors for science imaging or spectrography. From this analysis we identify key technologies, like cryogenic electronics, Ethernet based interfaces and fully-digital detectors, for implementing efficient control of many detectors. We also propose a concept joining all identified technologies that could be considered for future large ESO instruments as a complement of ESO's general detector controller, NGCII.
\end{abstract}

% Include a list of up to six keywords after the abstract

\begin{spacing}{2} % use double spacing for rest of manuscript

\newenvironment{conditions}
  {\par\vspace{\abovedisplayskip}\noindent\begin{tabular}{ll}}
  {\end{tabular}\par\vspace{\belowdisplayskip}\vspace{\belowdisplayskip}}

%% Introduction %% 
\section{Introduction}
\label{sect:intro}
With the European extremely large telescope (ELT) nearing completion and its first generation instruments under construction, proposals for future ESO {(European Southern Observatory)} telescopes and instruments are being called for{,} under the ``Expanding Horizons'' call for white papers. The aim of the call is to identify the next transformational facility that will advance humanity’s understanding of the {u}niverse{,} whilst fostering international collaboration. Such a transformational facility will require development in many areas of technology. 

From the detector engineering perspective a trend towards instruments using more and more detectors is observed. Hundreds of detectors will likely need to be considered in an instrument to meet science requirements. Besides challenging mechanical and optical design, one key problem identified in facilities of this scale is the implementation of a detector controller able to handle many detectors with reasonable cost, power budget, space and effort required for cabling.  

ESO’s current detector controller, the new general detector controller - second generation (NGCII), was designed to be extremely flexible and capable of operating a diverse selection of detectors in a variety of instruments. We first identify limitations of this modular design that make it less suitable for use in large instruments with many instances of the same detector. Then we perform a detailed study of existing large detector systems in astronomy using ten or more detectors with focus on systems using one hundred or more detectors and their detector controllers. From this rich body of experience from existing systems around the world, we identify some key technologies we consider important to future ESO detector controller development activities. 

Finally, we present a hypothetical detector controller that would implement the previously identified key technologies and provide a low-power compact solution to large scale detector control. We invite instrument consortia planning large scale instruments on ESO telescopes to enter a discussion on detector controllers early to achieve a cost-effective and sustainable solution. 

%%%%%%%%%%%%%%%%%%%%%%%%%%%%%%%%%%%%%%%%%%%%%%%%%%%%%%%%%%%%%
%%%%%%%                 Motivation                   %%%%%%%%
%%%%%%%%%%%%%%%%%%%%%%%%%%%%%%%%%%%%%%%%%%%%%%%%%%%%%%%%%%%%%
\section{Motivation}
\label{sec:motivation}
The current detector controller for use in all future instruments on ESO telescopes is NGCII\cite{Richerzhagen2024}. It is designed for testing and controlling any scientific detector suitable for ground-based astronomy and provides many features and options for in-system tuning during detector testing as well as extensive telemetry and signal mirroring functions to efficiently get new detectors up and running quickly and optimize the control parameters of individual detectors to operate the detectors at maximum performance. It has, however, some shortcomings regarding the control of large detector systems.

Due to the modularity of the NGCII controller size, cost, and power consumption will not scale well with number of detectors. NGCII generally allows control of one large ($4K\times4K$ or larger) or two small ($2K\times2K$) detectors per two height unit (2U) detector control hardware unit.

Besides those constraints, there are also limitations of the overall concept of a warm detector controller that can be placed a few meters ($<10m$) away from the cryostat. This concept means that many sensitive analog and digital signals need to be transmitted over that distance, which leads to complex and expensive cables and connectors. Also, this often implies the need {for} a cryogenic or warm preamplifier.

All effects listed above affect the economy and sustainability of large instruments{,} so to facilitate them on ESO telescopes we look beyond NGCII and investigate an option to provide a low-cost {and} highly integrated detector control solution.

%%%%%%%%%%%%%%%%%%%%%%%%%%%%%%%%%%%%%%%%%%%%%%%%%%%%%%%%%%%%%
%%%%%%%          Large Detector Systems              %%%%%%%%
%%%%%%%%%%%%%%%%%%%%%%%%%%%%%%%%%%%%%%%%%%%%%%%%%%%%%%%%%%%%%
\section{Review of Large Detector Systems}
\label{sec:previous_work}
Existing solutions to the problem of large detector system control are studied. Focus is put on detector systems in ground-based astronomy with 10 {or more} scientific imaging or spectrograph detectors. Space-based systems are considered if they show interesting features to be studied for ground-based astronomy. Instruments in other fields of science, like high-energy physics, are not considered, to keep the scope of the paper concise. 

%%%%%%%     Detector Control Model    %%%%%%%
\subsection{Detector Control Model}
\begin{figure}[ht]
\centering
\includegraphics[width=\textwidth]{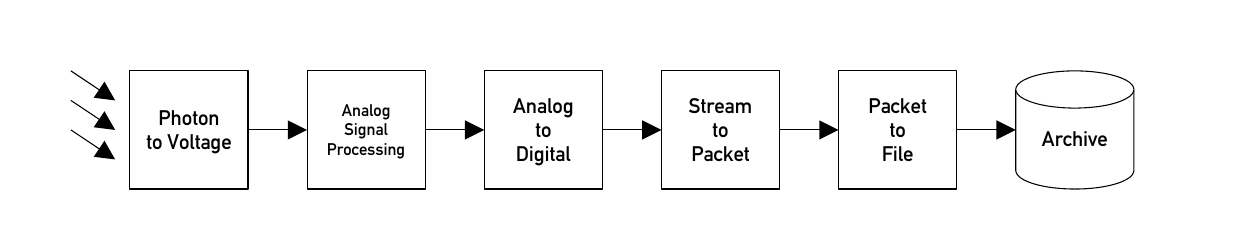}
\caption{Detector Control Model}
\label{pic:det_model}
\end{figure}

When analyzing existing solutions to the problem of reading out multiple detectors in ground-based astronomy, distinct concepts were implemented by various institutions. A basic detector control model shown in figure \ref{pic:det_model} is used to classify them. 

The model shows a flow diagram from photons to an archived science image along the video signal chain. For simplicity, the generation of control signals like clocks and biases is not included in the model. 

%%%%%%%     Large ESO Instruments     %%%%%%%
\subsection{Large ESO Instruments}
\label{subsec:eso_inst}
ESO telescopes are hosting several large instruments that meet the criteria {regarding application and number of detectors} outlined above. Since they all use very similar detector control schemes, they are described briefly. {OmegaCAM}\cite{Iwert2006} and VISTA Infrared Camera (VIRCAM) \cite{Sutherland2015} are based on the {FIERA (fast imager electronic readout assembly)} and {IRACE (infrared detector high speed array control and processing electronic)} controllers which are no longer comparable to modern detector control systems, so they are excluded from this study.

\begin{figure}[ht]
\centering
\includegraphics[width=\textwidth]{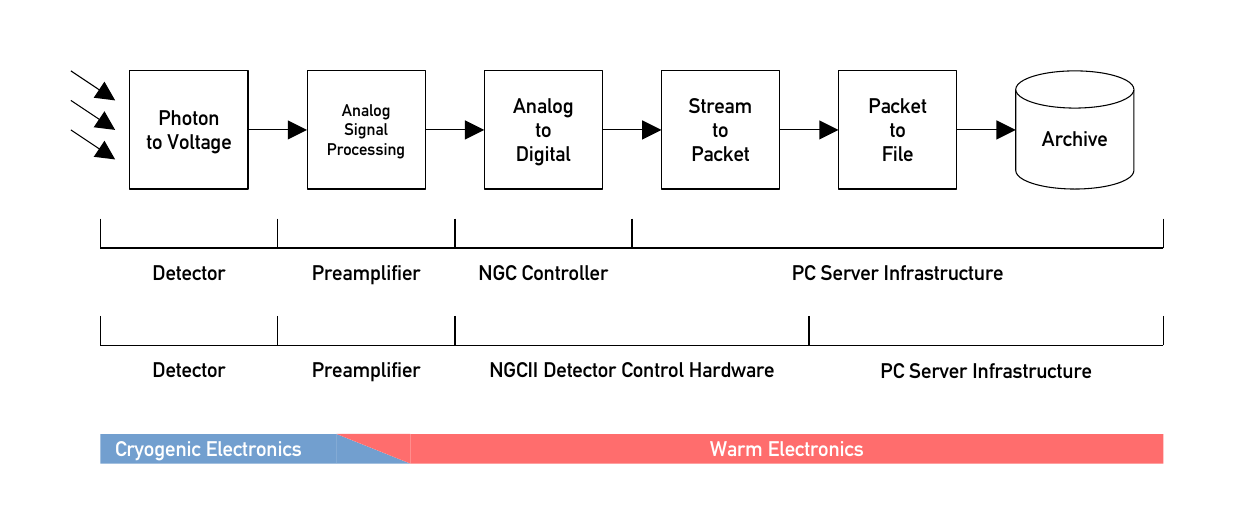}
\caption{Detector Control Model, NGC (top) and NGCII (bottom)}
\label{pic:det_model_eso}
\end{figure}

In general, ESO detector controllers consist of detector control hardware operated at telescope ambient temperature and located up to a few meters away from the cryostat. A cryogenic or warm analog preamplifier is generally present. Cryogenic preamplifier design at ESO is based on practical experience with a limited set of industrial grade components. 

{Both the new general detector controller (NGC) and NGCII are generic, modular detector controllers that are adapted to the connected detectors by populating the required modules in a common chassis. There is a subtle conceptual difference in how modularity is implemented. NGC groups all basic functions to read a detector on a single front-end basic module, adding video channel modules where necessary. NGCII modules mostly provide one function (clock, bias, video) with different module variants for CCD (charge coupled device) and CMOS (complementary metal-oxide-semiconductor) detectors.}

NGC streams image data to the PC server based detector workstation equipped with a custom {peripheral component interconnect (PCI) express} receiver card, using a custom fiber based link. The NGCII detector control hardware communicates with standard PC hardware based detector workstations over standard optical fiber Ethernet links.

\subsubsection{MUSE and BlueMUSE}
\label{subsec:eso_muse}
The MUSE {(Multi Unit Spectroscopic Explorer)} instrument\cite{Reiss2012} on the VLT uses 24 visible-wavelength $4K\times4K$ CCD231-84 detectors with {NGC}. {The instrument contains four detector controllers each controlling six detectors.}

For the proposed instrument BlueMUSE\cite{Richard2024}, which expands the capabilities of MUSE in the 350nm to 580nm wavelength range by adding sixteen $4K\times4K$ blue sensitive CCD detectors, {NGCII} is foreseen. {In current concepts the use of eight detector controller units each controlling two detectors is planned.}

\subsubsection{MICADO}
The MICADO {(multi-adaptive-optics imaging camera for deep observations)} instrument\cite{Sturm2024} currently under construction uses 9 infrared-wavelength $4K\times4K$ HAWAII-4RG-15 {(HgCdTe Astronomical Wide Area Infrared Imager)} detectors, cryogenic preamplifiers and NGCII detector controllers. While it has less than 10 detectors, it highlights both the advantages and {the} shortcomings of the ESO detector controller concept. NGCII supports infrared hybrid CMOS detectors by design due to its modular system, but with MICADO it reaches some of its limits outlined in section \ref{sec:motivation}. Eight of the MICADO detectors are operated using 32 video output channels while the central detector in the $3\times3$ mosaic uses 64 channels. There are ten warm detector cables, one per detector and an additional extension cable for the central detector, reaching a diameter of up to $36mm$ due to the amount of shielding required to facilitate a cable run of several meters in a noisy telescope environment.

\subsubsection{Analysis}
ESO detector controllers aim at controlling every suitable detector within and beyond the limits of its datasheet. For NGC and NGCII this is achieved by using a {generic,} modular system that can be adapted accordingly. The detector controller can be configured for each instrument with no, or minimal, development effort and does not require a high degree of integration into the instrument. Drawbacks of this approach are described in section \ref{sec:motivation} already.

{Despite the conceptual differences between NGC and NGCII described at the top of section \ref{subsec:eso_inst}, the two controllers are more similar to each other than to other detector control solutions studied in this paper. For example, MUSE and BlueMUSE, despite the differences in controller configuration described in section \ref{subsec:eso_muse}, both reach a packing density of approximately one controlled CCD231 per height-unit. In the context of this paper NGC and NGCII are interchangeable.}

%%%%%%%    Zwicky Transient Facility     %%%%%%%
\subsection{Zwicky Transient Facility (ZTF)}
\label{subsec:ztf}
The Zwicky Transient Facility (ZTF) uses 16 $6K\times6K$ CCD231-C6 science detectors in a $4\times4$ mosaic. Its design is described in Ref.\citenum{Dekany2020}. 

\begin{figure}[ht]
\centering
\includegraphics[width=\textwidth]{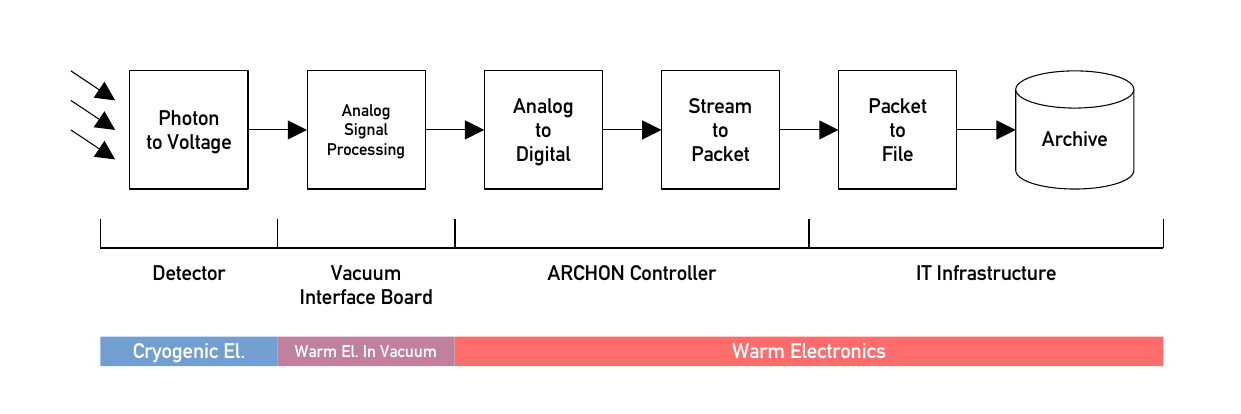}
\caption{Detector Control Model, ZTF}
\label{pic:det_model_ztf}
\end{figure}

ZTF implements the vacuum interface board (VIB), a printed circuit board {(PCB)} covering the entire focal plane, acting as vacuum {feedthrough} and housing the preamplifier electronics. The temperature of the VIB is not actively controlled but is found to equalize at approximately ambient temperature\cite{Dekany2020}. { Five instances of the STA (Semiconductor Technology Associates, Inc.)} Archon {generic, modular detector} controller\cite{Bredthauer2014,archon_manual} are used{,} mounted to the outside of the telescope tube. Archon communicates with the outside world via a Gigabit Ethernet interface\cite{Bredthauer2014,archon_manual}.

\subsubsection{Analysis}
Using a single PCB for electronics and vacuum {feedthrough} is an interesting approach that reduces the number of parts required, but it is hard-limited by the size of PCB panel that can be manufactured. The designers identified a $24'' \times 18''$ panel size limit (approximately $600mm \times 450mm$)\cite{Dekany2020}, so for larger focal planes the concept becomes unfeasible. Furthermore, a single defect requires the repair or replacement of the entire PCB. Acceptability of this approach needs to be carefully analyzed for future projects. 

{The STA Archon detector controller\cite{Bredthauer2014,archon_manual} is similar to NGCII in that it is a generic and modular controller, providing dedicated modules per function (clock, bias, video).  It is more compact than the ESO detector controllers and primarily targets CCD detectors. Multiple detectors can be controlled from a single chassis.} The designers of the instrument worked around the pre-determined form-factor of the {commercial off-the-shelf} detector controllers by placing them outside the telescope tube requiring water cooling of each individual unit\cite{Dekany2020}.

%%%%%%%   HETDEX VIRUS   %%%%%%%
\subsection{HETDEX / VIRUS}
\label{subsec:virus}
The VIRUS instrument for the Hobby-Eberly Telescope Dark Energy Experiment (HETDEX) uses 78 spectrograph detector heads containing two detectors each\cite{Hill2021}. There is a total of 156 STA3600 CCD detectors with $2K \times 2K$ resolution each, custom-made by STA. One Astronomical Research Cameras, Inc. (ARC) controller, located in a warm enclosure central to the detector head in-between the two detectors, is used, reading out the CCD video output in two-channel mode\cite{Hill2021,Hill2012}. 

\begin{figure}[ht]
\centering
\includegraphics[width=\textwidth]{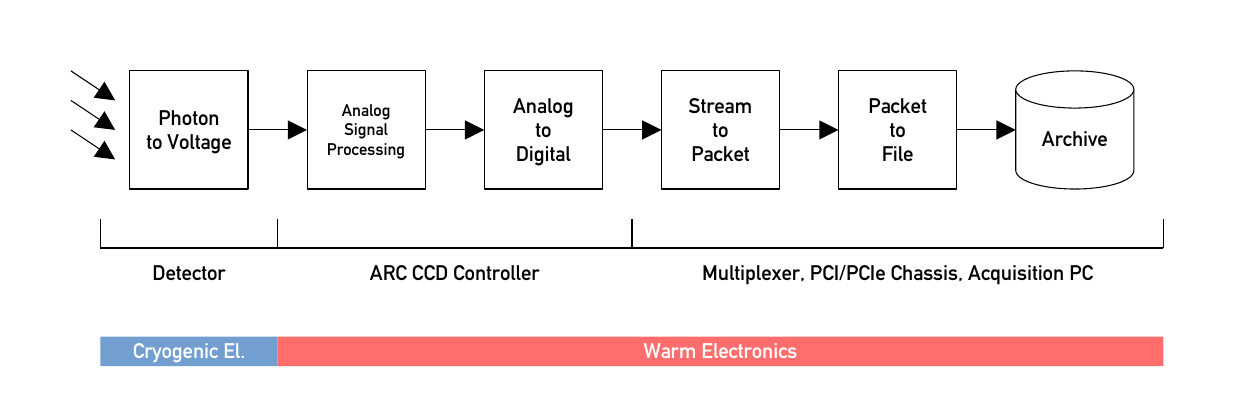}
\caption{Detector Control Model, HETDEX}
\label{pic:det_model_hetdex}
\end{figure}

The hardware generation and configuration of the ARC controller used by VIRUS is not specified in literature. Due to the dissolution of Astronomical Research Cameras, Inc., official documentation is no longer publicly available online but unrelated publications\cite{Leach2000} describe the general architecture of the controller. The controller is modular and uses a proprietary fiber uplink to connect to a PCI or {PCI express (PCIe)} interface board populated in a PC system. The VIRUS instrument uses a custom 8:1 multiplexer to combine eight data links into one\cite{Hill2012}. The output of each multiplexer is connected to a PCI card, populated in a PCI to PCIe expansion chassis, interfacing the VIRUS Data Acquisition System (VDAS) PC\cite{Hill2012}.

\subsubsection{Analysis}
Due to its nature as a fiber-fed spectrograph, the instrument does not use a single focal plane, but {many} dual detector heads mounted in a shelf-like construction. This breaks down the detector control problem into the $78\times$ replication of a relatively small detector controller with short cable runs, located very close to the detectors. The relatively large and accessible system allows easy installation in less than 3 hours per unit\cite{Spencer2018}, and thus easy maintenance, but does impose a challenge for cooling of two very large structures inside the telescope enclosure\cite{Spencer2018,Hill2021}.

The designers developed a custom multiplexer\cite{Hill2012}, reducing the number up links to the data acquisition PC. This illustrates an issue using proprietary links, that any equipment interfacing them needs to be custom-made while for standard links, like Ethernet, commercial equipment is available.

%%%%%%%   DESI   %%%%%%%
{
\subsection{Dark Energy Spectroscopic Instrument}
\label{subsec:desi}
The dark energy spectroscopic instrument (DESI) is a fiber-fed spectrograph on the Mayall telescope in Arizona. It uses 10 spectrograph heads with three detectors each. The spectrograph heads are located in a temperature controlled shelf-like structure\cite{Abareshi2022}. Of the three $4K \times 4K$ CCD detectors per spectrograph, the blue channel is manufactured by STA while the red and near-infrared channel CCD design originates at the Lawrence Berkeley National Laboratory (LBNL)\cite{Martini2018}. Each of the thirty detectors is located in its own dedicated cryostat. No cryogenic electronics are mentioned in literature but a custom flex circuit to break out the detector signals to a connector. LBNL also designed the custom, warm detector controller common to all three channels. The controller is mounted externally to the cryostat\cite{Edelstein2018}.

\begin{figure}[ht]
\centering
\includegraphics[width=\textwidth]{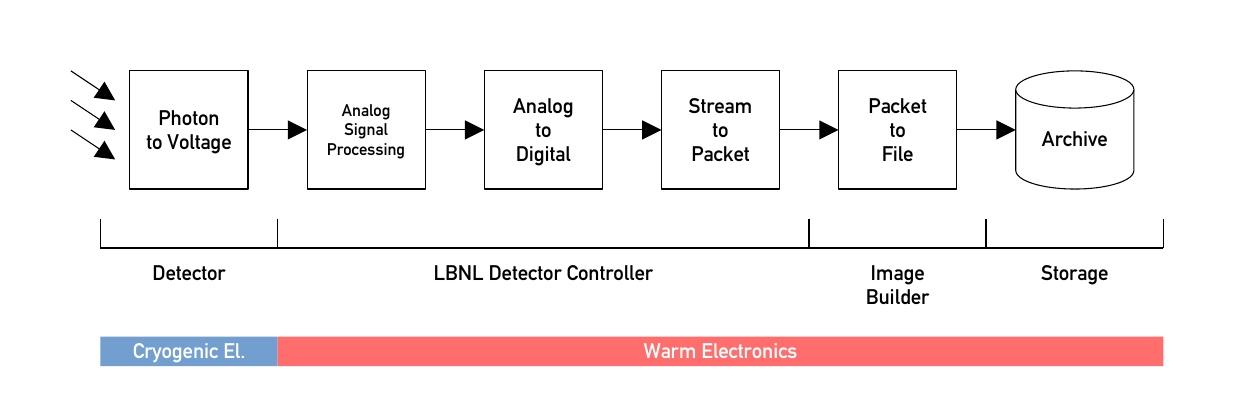}
\caption{Detector Control Model, DESI}
\label{pic:det_model_desi}
\end{figure}

The detector controller receives commands and transmits image data via Gigabit Ethernet\cite{Edelstein2018} to the PC based image builder system that assembles FITS files and forwards them to storage via another Gigabit Ethernet link\cite{Martini2018}.

\subsubsection{Analysis}
With a very short distance between detector and controller, entirely warm electronics, standard Ethernet links and PC hardware DESI has found a very simple detector control solution. The designers highlight the fact that they are able to control two types of CCD (STA and LBNL) with the same model of detector controller\cite{Edelstein2018}. This avoids maintaining two types of controller in the long term.

Like HETDEX/VIRUS described in section \ref{subsec:virus}, DESI is a fiber-fed spectrograph. Compared to a focal plane system there are different challenges for the detector controller. While build volume and accessibility of the system are less problematic, the compact arrangement of detectors in a focal plane can be advantageous when it is possible to control multiple detectors with one controller.

A major difference between HETDEX/VIRUS and DESI is the location of the detector heads. While HETDEX/VIRUS places them in the telescope environment (also see section \ref{subsec:virus}), DESI places them in a temperature controlled and vibration dampened enclosure in the Coud\'e room of the telescope. If the location of the detector controller is known early in the design process, the specific environmental conditions and interactions with other subsystems can be taken into account and used to simplify controller design.
}

%%%%%%%    LSST Camera    %%%%%%%
\subsection{LSST Camera}
\label{subsec:lsst}
The LSST Camera is the only instrument of the former Large Synoptic Survey Telescope (LSST), now Vera C. Rubin Observatory. For scientific imaging it uses 189 custom $4K \times 4K$ CCD detectors\cite{Kahn2010,OConnor2012}. They are organized into 21 raft towers of nine detectors each. Read-out electronics are closely integrated into the raft tower, split into two distinct assemblies.

\begin{figure}[ht]
\centering
\includegraphics[width=\textwidth]{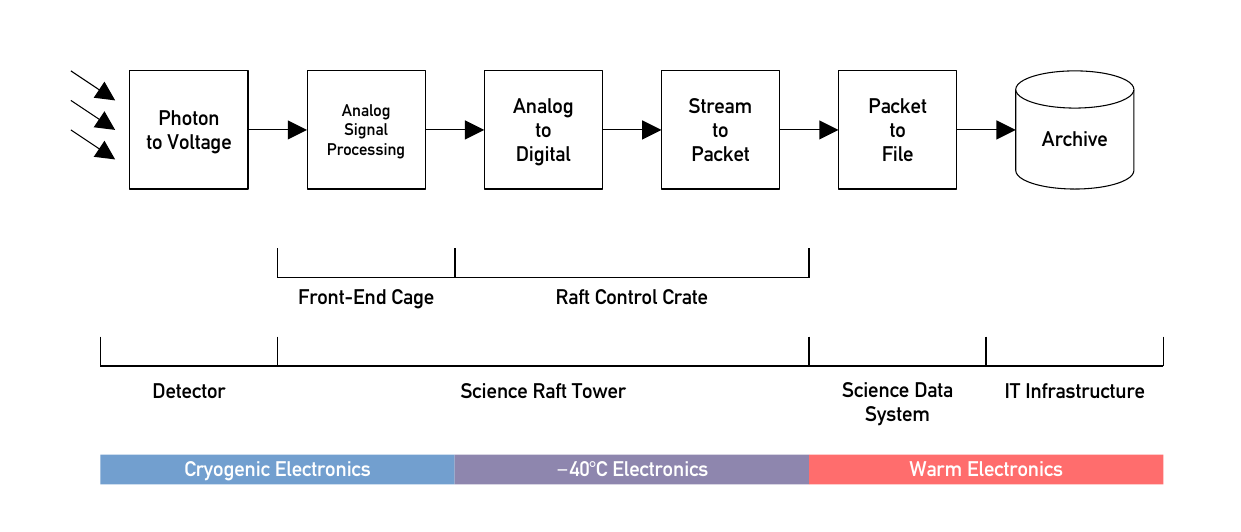}
\caption{Detector Control Model, LSST Camera}
\label{pic:det_model_lsst}
\end{figure}

The front-end cage (FEC)  contains boards operate at around $-100^\circ C$, like the detectors. They contain two types of custom {application-specific integrated circuits (ASICs)} for detector control\cite{OConnor2012}, {ASPIC (analog signal processing ASIC)} implementing a dual slope integrator and {CABAC (clocks and biases ASIC for CCDs)}, generating detector control signals.

The raft control crate (RCC) is temperature controlled to $-40^\circ C$, which allows the use of commercial components within their specified temperature range\cite{OConnor2012}, saving qualification effort. It contains the analog to digital conversion circuit as well as a {field programmable gate array (FPGA)} that packetizes the video signal data and transmits it on a high speed serial link to the external science data system. { No specification of the physical layer (optical/electrical) was found in literature.}

The science data system (SDS) is based on the advanced telecommunications computing architecture (ATCA)\cite{Atca}. It is adapted from the Stanford Linear Accelerators (SLAC) reconfigurable cluster element\cite{Herbst2014} (RCE) platform including $60 TB$ of archival storage\cite{Kahn2010}.

\subsubsection{Analysis}
The LSST camera detector control system is an impressive implementation of 189 butted detectors in a single focal plane. The designers managed to place all detector control electronics behind the focal plane. The division of the focal plane into raft towers allows stand-alone testing and maintenance of parts of the focal plane. This deep integration of electronics with the detectors is facilitated by two-tier thermal management, transitioning to commercially available electronics as early in the signal chain as possible. 

The large scale of the instrument allowed the designers to choose the development of two custom ASICs for the electronics{,} not possible to implement at $-40^\circ C$. Among the instruments studied in this paper, custom development of ASICs for ground-based astronomy is unique. Cost and long development timelines are normally prohibitive.

Another notable feature of the LSST Camera is the adoption of hardware and concepts from the high-energy physics field with the ATCA based science data system and data link between RCE and SDS. To a limited extent ESO's NGCII follows a similar approach using {the micro telecommunications computing architecture (MicroTCA)}, originally developed in the high-energy physics community. This suggests that for future large instruments a cross-disciplinary collaboration should be considered if promising concepts are available.

Finally, {the} LSST Camera uses a different approach to all other designs studied regarding data processing. The use of an industrial ATCA based processing system including limited storage, instead of standard PC hardware, is unique to the LSST camera.

%%%%%%%   Space Based ASIC Systems   %%%%%%%
\subsection{Space Based ASIC Systems}
\label{subsec:space_asic}
There exists a class of systems based on analog detectors and dedicated ASICs for readout frequently used in space applications which is discussed {here} as a whole. While not originally intended for the use in ground-based astronomy, these systems can be adapted to the use case.

\begin{figure}[ht]
\centering
\includegraphics[width=\textwidth]{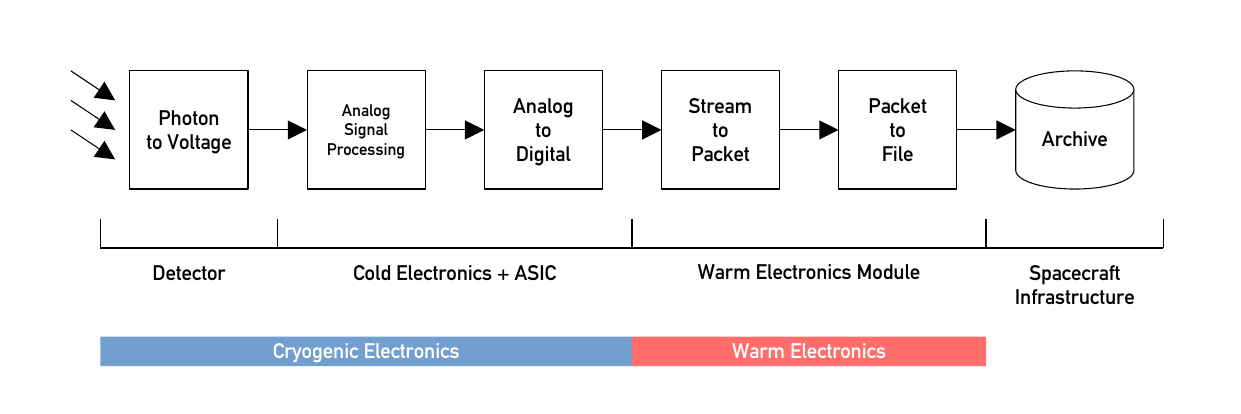}
\caption{Detector Control Model, ASIC}
\label{pic:det_model_asic}
\end{figure}

The {Euclid} space missions {near-infrared spectrometer and photometer (NISP)}, uses 16 Teledyne Imaging Sensors $2K\times2K$ HAWAII-2RG infrared hybrid CMOS detectors and the SIDECAR {(system for image digitization, enhancement, control and retrieval)} ASIC\cite{Loose2005} designed as companion part by the same company. The SIDECAR ASIC integrates all analog electronics for detector control and video {analog-to-digital converters (ADC)} for sampling the analog video outputs of the detector into a single chip. 

The Wide-Field Instrument (WFI) is the primary instrument of the Nancy Grace Roman Telescope. It uses 18 Teledyne Imaging Sensors $4K\times4K$ HAWAII-4RG{-10} infrared hybrid CMOS detectors\cite{Mosby2020} paired with {ACADIA\cite{Loose2018} (ASIC for control and digitization of imagers for astronomy) } specifically designed for the project.

Ref.\citenum{Loose2012} presents a warm control electronics concept for a mosaic of 32 HAWAII-xRG detectors with SIDECAR. Possible ground-based instrument applications for the ELT, Thirty Meter Telescope (TMT) and Giant Magellan Telescope (GMT) are explicitly mentioned. The concept proposes to mount ASIC control cards and an active back plane close to the cryostat and connect the assembly to a PC through an optical camera link connection and a frame grabber.

The Gaia space telescope uses 106 $4K\times2K$ CCD detectors in a single focal plane. The design is summarized from Ref.\citenum{Gaia2016}. While the instrument uses three detector types for different wavelengths, they are based on the same architecture and driven by identical electronics. Each detector is connected to a dedicated proximity electronics module (PEM) containing control electronics as well as the ADCs, with radiated cooling to space. Groups of PEMs, organized by focal plane row, are connected to video processing units (VPU) running the video processing algorithm. The Gaia detector control system is a dedicated design for space operation and has limited applicability for ground-based astronomy. Hyper Suprime-Cam described in section \ref{subsec:subaru} is a more relevant example of a comparable system.

\subsubsection{Analysis}
With space-based systems having their own set of challenges, applicability to ground-based astronomy is limited. Ground-based instruments have technologies available that cannot be used on space-qualified hardware. The concept described in Ref.\citenum{Loose2012} is unique in proposing the use of ASICs for the space sector on the ground. The choice of a relatively simple digital interface, Camera Link, as opposed to the packet based Ethernet links use{d} elsewhere could be an indication of a different approach in that field as well.

A notable feature of the Gaia space telescope\cite{Gaia2016} is the use of mechanically and electrically similar detectors in three sensitivity configurations. This concept allows covering a wide wavelength range while using identical control electronics{,} reducing the effort required for development and testing. 

%%%%%%%   Subaru Telescope   %%%%%%%
\subsection{Hyper Suprime-Cam}
\label{subsec:subaru}
The Hyper Suprime-Cam wide-field imaging instrument on the Subaru Telescope uses 116 $2K\times4K$ CCD detectors in a single focal plane, 104 of which used for scientific imaging\cite{Miyazaki2018,Nakaya2012,Komiyama2017}.

\begin{figure}[ht]
\centering
\includegraphics[width=\textwidth]{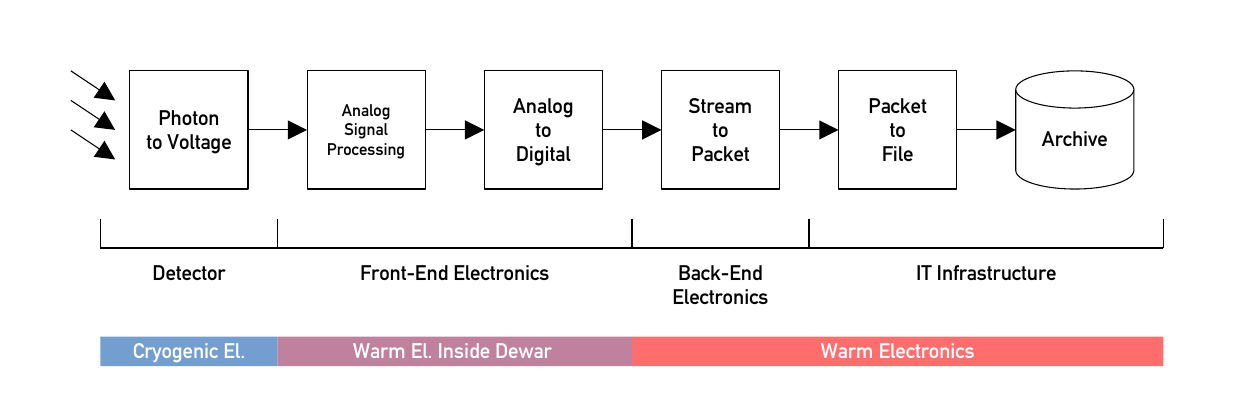}
\caption{Detector Control Model, Hyper Suprime-Cam}
\label{pic:det_model_subaru}
\end{figure}

Hyper Suprime-Cam integrates all analog electronics, including the analog to digital converters, onto front-end electronics modules integrated into the vacuum dewar but operated at near room temperature\cite{Komiyama2017}. The authors mention that heat transfer from electronics to ambient through structural parts is problematic but solve this issue by providing a passive thermal link and rating the front-end modules for operation at $+50^\circ C$. Aluminum core PCBs are used for thermal management\cite{Nakaya2012}. The vacuum vessel {feedthrough} carries digital LVDS {(low-voltage differential signaling)} signals to the back-end electronics which convert them to optical 1 Gigabit Ethernet links connecting to PCs in the observation room\cite{Miyazaki2018}.

\subsubsection{Analysis}
Hyper Suprime-Cam integrates warm electronics into the cryostat. While generally similar to the approach chosen by ZTF, here the warm electronics are much more modular. The designers reference a heritage of the detector system from older instruments\cite{Nakaya2012}. The papers studied show a lot of attention to detail regarding thermal management of warm electronics in vacuum.

%%%%%%%    Pan-STARRS    %%%%%%%
\subsection{Pan-STARRS}
\label{subsec:pan_starrs}

The Panoramic Survey Telescope and Rapid Response System (Pan-STARRS) is a survey instrument at Haleakala Observatory, Hawaii. Its camera uses 60 orthogonal transfer array CCDs specifically developed for the instrument. The design is summarized from Ref.\citenum{Onaka2008}. 

\begin{figure}[ht]
\centering
\includegraphics[width=\textwidth]{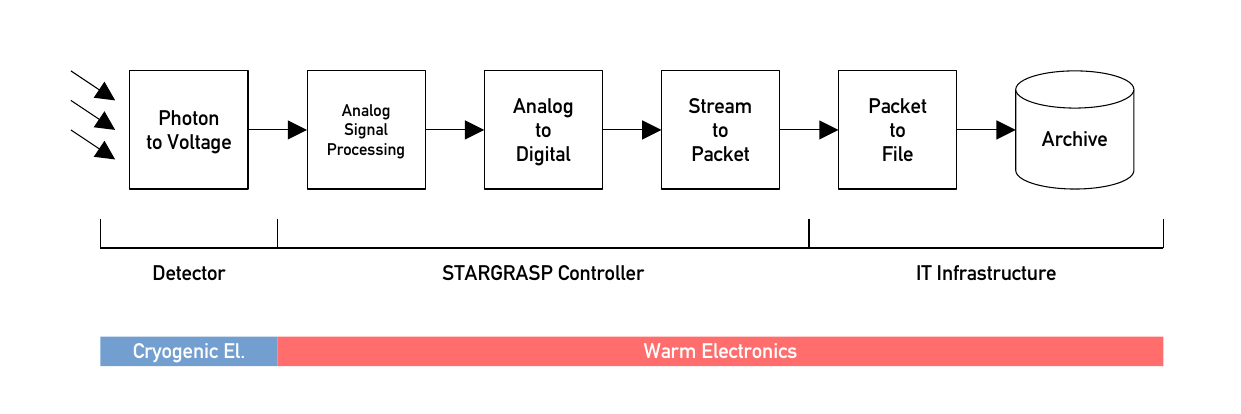}
\caption{Detector Control Model, Pan-STARRS}
\label{pic:det_model_pan_starrs}
\end{figure}

Due to the large number of 4500 control signals entering and exiting the instrument cryostat, and the need for real-time readout of parts of the array, it was chosen to use rigid-flexible PCBs for routing all signals to warm STARGRASP {(scalar topology architecture of redundant gigabit readout array signal processors)} detector controllers mounted horizontally next to the focal plane. These detector controller units communicate with pixel servers via Gigabit Ethernet links over standard network infrastructure components.

\subsubsection{Analysis}
While the approach of using warm detector controller electronics close to the focal plane appears common at first sight, the unique feature is the exclusive use of continuous rigid-flex printed circuit boards for internal wiring, vacuum {feedthrough} and a very short warm cable run. This avoids using cable assemblies and possibly expensive and bulky connectors. At the same time the chosen approach is not affected by the PCB panel size limit discussed in section \ref{subsec:ztf}. Since the rigid-flex PCBs are passive, there are no maintenance concerns and the warm control electronics are modular and can be changed quickly. The overall design is possible because the designers were able to use valuable space next to the focal plane for control electronics.

The designers mention that issues with cross-talk between detector signals were observed, despite their efforts to shield them from each other on the rigid-flex PCB assemblies, but ultimately fixed in software by tuning phase delays\cite{Onaka2008}. Each rigid-flex assembly is driven by a monolithic STARGRASP controller unit, and thus the same FPGA, enabling this tuning step. This outlines that synchronization of detector controllers on the same focal plane is a feature to be considered to combat expected or unexpected cross-talk issues. 

%%%%%%%   JPCam   %%%%%%%
\subsection{JPCam}
\label{subsec:jpcam}
JPCam is the instrument to perform the Javalambre Physics of the Accelerating Universe Astrophysical Survey (J-PAS)\cite{Bryant2024}. It uses 14 $9K\times9K$ CCD290-99 science detectors in a commercially manufactured camera described in Ref.\citenum{Jorden2012}.

\begin{figure}[ht]
\centering
\includegraphics[width=\textwidth]{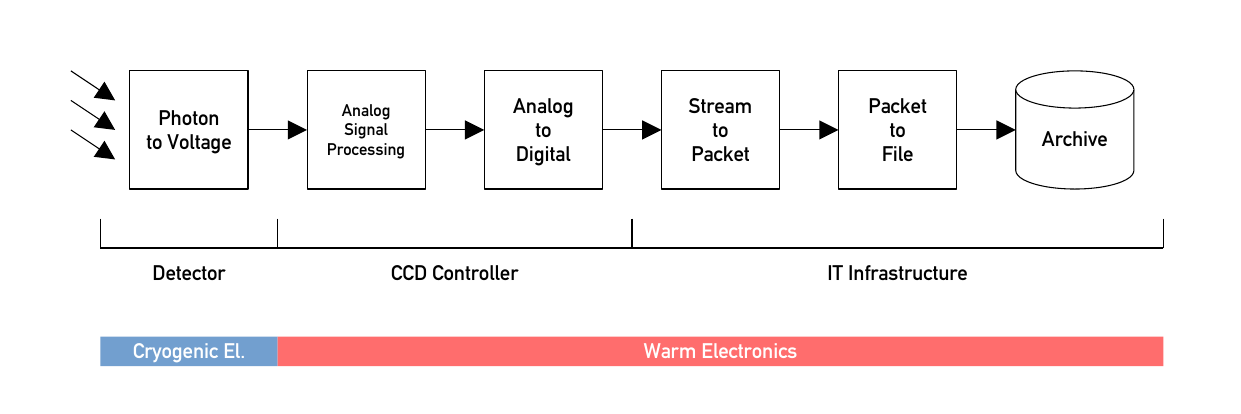}
\caption{Detector Control Model, JPCam}
\label{pic:det_model_jpcam}
\end{figure}

Detector signals exit the cryostat through individual flexible PCBs and connect to the CCD controller mounted behind the focal plane. The detector controller streams its pixel data to host PCs equipped with camera link frame grabbers over optical camera-link interfaces\cite{Jorden2012}.

\subsubsection{Analysis}
While the design of detector control is quite conventional, using warm electronics behind the focal plane and a connector-based vacuum {feedthrough}, the unique element of the design is that the entire camera was developed and produced commercially by Teledyne e2v, the manufacturer of the CCD detectors used. If system level requirements and interface definitions are clear, this approach can be very successful as demonstrated by J-PAS. 

%%%%%%%   TAOS-II   %%%%%%%
\subsection{TAOS-II}
\label{subsec:taosii}
The Transneptunian Automated Occultation Survey (TAOS-II)\cite{Lehner2012} at the Observatorio Astronomico Nacional in Baja California, Mexico uses three telescopes with identical cameras. Each camera consists of a mosaic of ten $4.5K\times2K$ Teledyne e2v CIS113\cite{Pratlong2016} CMOS detectors.

\begin{figure}[ht]
\centering
\includegraphics[width=\textwidth]{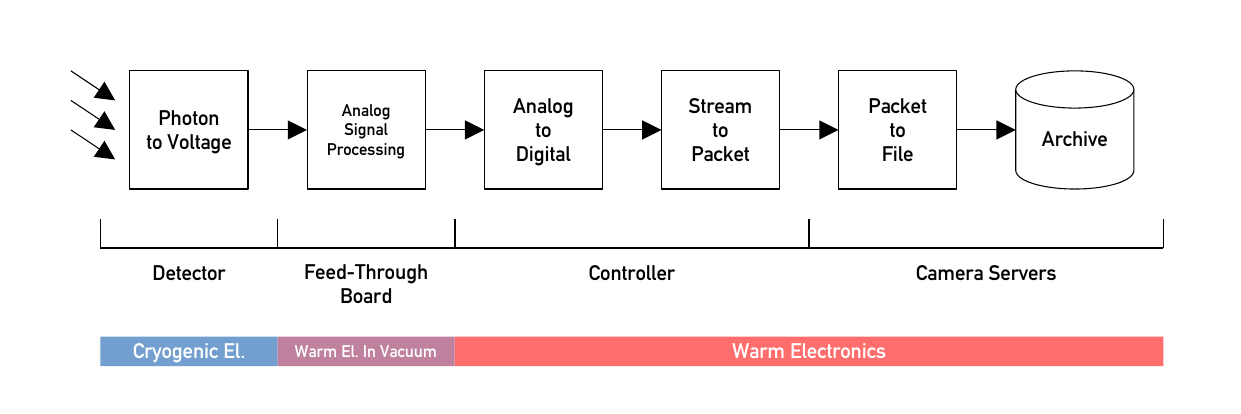}
\caption{Detector Control Model, TAOS-II}
\label{pic:det_model_taosii}
\end{figure}

Detector signals are routed out of the cryostat through two large printed circuit boards per camera containing preamplifier circuits. Video processor boards for each detector, carrying FPGA based piggyback boards, are plugged into the {feedthrough} PCB\cite{Wang2022}. The link between the Xilinx Kintex 7 FPGA boards and the camera servers \cite{Castro2021} is a standard Ethernet link\cite{Wang2022,Wang2014,Amato2012}. The designers use a dedicated optical fiber based system to synchronize the cameras distributed over three telescopes\cite{Wang2022}.

\subsubsection{Analysis}
The approach for the TAOS-II detector controller is similar to that of Pan-STARRS discussed in section \ref{subsec:pan_starrs} and the ZTF discussed in section \ref{subsec:ztf}. Like on ZTF, a printed circuit board with active preamplifier circuits is used but like on Pan-STARRS the detector control electronics are directly docked to the {feedthrough} PCB.

Using active electronics on large vacuum PCBs needs to be carefully considered for future large instruments evaluating maintenance concerns as already discussed in section \ref{subsec:ztf}. Again, the designers were able to use valuable space directly at the cryostat for detector control purposes yielding a system with no cable assemblies and all detector signals routed through printed circuit boards.

\subsection{Summary and Conclusion}
A summary of detector systems discussed in this section is given in table \ref{tab:overview}. It is concluded that most systems discussed use CCD detectors. The notable exceptions are the concept discussed in section \ref{subsec:space_asic} adapting space based infrared CMOS detectors with companion ASIC to ground-based astronomy and TAOS-II, discussed in section \ref{subsec:taosii} using visible CMOS detectors.

For the detector controller there are options with various degrees of integration. From the LSST Camera which integrates cold electronics including an FPGA into the cold domain, over other concepts solving the issue of bringing parts of the detector control electronics into the vacuum vessel to concepts like JPCam and HETDEX / VIRUS that use a full warm detector controller located close to the detector vessel containing only the detector.

\begin{table}[ht]
\caption{System Summary} 
\label{tab:overview}
\scriptsize
\begin{center}
\begin{tabular}{|l|l|l|p{0.3\textwidth}|p{0.2\textwidth}|}
\hline
\rule[-1ex]{0pt}{3.5ex} Ch.                       
                        & Instrument/System   
                        & Detector
                        & Signal Chain
                        & Comm. Link \\ \hline \hline
\rule[-1ex]{0pt}{3.5ex} \ref{subsec:eso_inst}     
                        & Various NGC   
                        & Various CCD and CMOS
                        & Cold or warm preamplifier. Warm detector control hardware. PC based detector workstation.
                        & Custom Opt. Link \\ \hline
\rule[-1ex]{0pt}{3.5ex} \ref{subsec:eso_inst}     
                        & Various NGCII\cite{Richerzhagen2024} 
                        & Various CCD and CMOS
                        & Cold or warm preamplifier. Warm detector control hardware. PC based detector workstation.
                        & Ethernet 1GbE/10GbE \\ \hline
\rule[-1ex]{0pt}{3.5ex} \ref{subsec:ztf}          
                        & ZTF - Archon\cite{Bredthauer2014} 
                        & $16\times$ CCD231-C6
                        & Full focal plane {feedthrough} PCB with integrated warm vacuum preamp and Archon controller mounted to telescope tube.
                        & Ethernet 1GbE \\ \hline
\rule[-1ex]{0pt}{3.5ex} \ref{subsec:virus}        
                        & HETDEX - VIRUS\cite{Hill2021} 
                        & $156\times$ STA3600 CCD
                        & Two vacuum vessels per detector head including warm ARC controller. Custom uplink multiplexer and video data acquisition PC with PCI/PCIe interface cards.
                        & Custom Opt. Link \\ \hline
\rule[-1ex]{0pt}{3.5ex} \ref{subsec:desi}        
                        & {DESI\cite{Abareshi2022}}
                        & $30\times$ CCD
                        & One cryostat per detector with a dedicated warm detector controller each. PC based image builder connected via Ethernet.
                        & Ethernet 1GbE \\ \hline
\rule[-1ex]{0pt}{3.5ex} \ref{subsec:lsst}
                        & LSST Camera\cite{OConnor2012}
                        & $189\times$ Custom CCD
                        & Nine-detector raft tower containing cryogenic front-end {cage} with CABAC and ASPIC ASICs and $-40^\circ C$ raft control crate. ATCA based science data system.
                        & SLAC PGP\cite{Herbst2014} \newline Ethernet 10GbE\cite{Kahn2010}\\ \hline
\rule[-1ex]{0pt}{3.5ex} \ref{subsec:space_asic}   
                        & Markury Scientific Demo\cite{Loose2012} 
                        & $32\times$ HAWAII-xRG 
                        & SIDECAR ASIC and warm data electronics module.
                        & Opt. Camera Link \\ \hline
\rule[-1ex]{0pt}{3.5ex} \ref{subsec:subaru}       
                        & Hyper Suprime-Cam\cite{Miyazaki2018} 
                        & $104\times$ CCD
                        & Modular warm front-end electronics in vacuum vessel. Digital interface with warm back-end electronics.
                        & Ethernet 1GbE \\ \hline
\rule[-1ex]{0pt}{3.5ex} \ref{subsec:pan_starrs}   
                        & Pan STARRS\cite{Onaka2008} 
                        & $60\times$ OTA CCD
                        & Passive rigid-flex printed circuit board vacuum {feedthrough} with directly attached warm detector controller modules.
                        & Ethernet 1GbE \\ \hline
\rule[-1ex]{0pt}{3.5ex} \ref{subsec:jpcam}        
                        & JPCam\cite{Jorden2012} 
                        & $14\times$ CCD290-99
                        & Commercially produced detector system with warm detector controller mounted behind focal plane.
                        & Opt. Camera Link \\ \hline
\rule[-1ex]{0pt}{3.5ex} \ref{subsec:taosii}        
                        & TAOS-II\cite{Lehner2012} 
                        & $30\times$ CIS113
                        & Printed circuit board with active preamplifier as vacuum {feedthrough}. Detector control electronics plugged into {feedthrough}.
                        & Ethernet 1GbE\cite{Wang2022,Wang2014,Amato2012} \\ \hline
\end{tabular}
\end{center}
\end{table} 

At least at one point in the system every detector control system uses a high-speed link to bridge physical distance. Most use Ethernet while others rely on custom optical links or optical Camera Link. All systems use standard PC hardware / IT infrastructure for image data processing except for the LSST camera that processed data on an ATCA based system close to the telescope.

%%%%%%%%%%%%%%%%%%%%%%%%%%%%%%%%%%%%%%%%%%%%%%%%%%%%%%%%%%%%%
%%%%%%%                  Technologies                %%%%%%%%
%%%%%%%%%%%%%%%%%%%%%%%%%%%%%%%%%%%%%%%%%%%%%%%%%%%%%%%%%%%%%
\section{Technologies and Constraints}
\label{sec:technologies}
After analyzing existing systems in section \ref{sec:previous_work} and with the motivation outlined in section \ref{sec:motivation} we identify key technologies to be investigated for future large detector systems as ESO.

%%%%%%%  ASICs and FDD %%%%%%%
\subsection{ASICs and Fully Digital CMOS Detectors}
From the instruments studied in section \ref{sec:previous_work} all those instruments operating electronics at cryogenic temperatures use ASICs in their signal processing chain. LSST uses ASPIC and CABAC for visible wavelength CCD control\cite{OConnor2012}. For the HAWAII-xRG hybrid infrared detectors SIDECAR\cite{Loose2012} and ACADIA\cite{Loose2018} are available. If performance of an existing ASIC meets a new instruments requirements, and the ASIC is available for purchase, it should be given consideration. 

Development of a new ASIC is costly and needs to be weighed against other factors. {LSST has shown that developing a project-specific ASIC for an instrument using many similar or identical detectors can be successful. Due to the time-scales involved with custom ASIC development and testing, it is important to have stable detector requirements early. If science requirements allow, it is advantageous to avoid heterogenous instrument architectures increasing the ratio of silicon produced versus development effort. If the number of chips required to warrant a custom ASIC design is not met by a single instrument, a joint development of a generic ASIC for a family of detectors used by several instruments can be considered.} 

\begin{table}[ht]
\caption{List of ASICs in Active Deployment} 
\label{tab:asics}
\small
\begin{center}
\begin{tabular}{|l|l|l|l|l|l|}
\hline
\rule[-1ex]{0pt}{3.5ex} ASIC & Detector & Type & Interface & Ref.\\ \hline \hline
\rule[-1ex]{0pt}{3.5ex} SIDECAR & HAWAII-xRG CMOS & Digital with ADC & LVDS or CMOS & \citenum{Loose2003} \\ \hline
\rule[-1ex]{0pt}{3.5ex} ACADIA & HAWAII-xRG CMOS & Digital with ADC & LVDS & \citenum{Loose2018} \\ \hline
\rule[-1ex]{0pt}{3.5ex} ASPIC + CABAC & CCD & Analog & - & \citenum{Antilogus2017} \\ \hline 
\end{tabular}
\end{center}
\end{table} 

Recently, more and more fully-digital detectors are entering the market. Fully-digital CMOS detectors integrate the analog to digital converters and some bias generation circuitry on the detector or read-out IC die. ESO-internal studies for future instruments have brought up the detectors in table \ref{tab:detectors} as possible candidates for further evaluation with their electro-optical properties suitable for ground-based astronomy. The {GeoSnap-18 read-out integrated circuit (ROIC)} is successfully run at ESO with the NGCII detector controller\cite{Richerzhagen2025}.

\begin{table}[ht]
\caption{List of Fully-Digital Detectors} 
\label{tab:detectors}
\footnotesize
\begin{center}
\begin{tabular}{|l|l|l|l|l|r|l|}
\hline
\rule[-1ex]{0pt}{3.5ex} Manufacturer & Detector & Band & Pixels & Interface & Pwr. &  Ref.\\ \hline \hline
\rule[-1ex]{0pt}{3.5ex} GPIXEL & GSense 6060BSI & Vis. & 6144x6144, 10\textmu m & 50x LVDS @ 630Mbps & 5.0W &  \citenum{gsense6060}\\ \hline
\rule[-1ex]{0pt}{3.5ex} GPIXEL & GSense 1081BSI & Vis. & 8900x9120, 10\textmu m & 5x LVDS @ 250Mbps & 1.4W &  \citenum{gsense1081}\\ \hline
\rule[-1ex]{0pt}{3.5ex} GPIXEL & GSense 1517BSI & Vis. & 4116x4100, 15\textmu m & 10x LVDS @ 420Mbps & 1.0W & \citenum{gsense1517}\\ \hline 
\rule[-1ex]{0pt}{3.5ex} Teledyne E2V & CIS30x-66 & Vis. & 9000x8300, 10\textmu m & 12x CML @2Gbps & 4.8W & \citenum{cis30x}\\ \hline
\rule[-1ex]{0pt}{3.5ex} Teledyne TIS & {GeoSnap-18} & IR & 2048x2048, 18\textmu m  & 8x CML @1.6Gbps & 0.9W & \citenum{geosnapFPA}\\ \hline
\end{tabular}
\end{center}
\end{table} 

When discussing fully digital detectors in table \ref{tab:detectors} it was noted that there are two distinct interface types for transmitting video signals. This is also partially applicable to the ASICs in table \ref{tab:asics} where SIDECAR and ACADIA implement a similar interface.

Synchronous {LVDS} interfaces use 8 to 50 LVDS data outputs and one double data rate (DDR) clock output used by the receiver to sample the data signals. The recommended maximum data rate by the LVDS electrical standard ANSI/TIA/EIA-644\cite{tia_644a_2001} is 655Mbps, but faster data rates can be used if the entire transmission channel is under control of the implementer. LVDS is a voltage mode standard and does not implement measures to improve signal integrity such as drive strength adjustment or pre-emphasis. LVDS signals are received by differential, general purpose inputs on the controller FPGA or {system-on-chip (SoC)}.

Other detectors implement a current mode logic (CML) interface with 8b/10b coded data and embedded clocking. These operate at up to 2Gbps and require the connection to a multi-gigabit receiver on the detector controller FPGA/SoC.

%%%%%%%  Thermal Limits for Electronics %%%%%%%
\subsection{Thermal Limits for Electronics}
\label{subsec:thermal_limits}
Low temperature operation of electronics is an important aspect to study for the concept at hand. Several component classes are affected by low temperature operation which needs to be considered for design. 

The LSST camera design as summarized in section \ref{subsec:lsst} makes the conscious choice to operate part of their integrated electronics, the raft control crate (RCC), at $-40^\circ C$ to use standard commercial electronics versus their front-end cage (FEC) operating at $-100^\circ C$. Systems with a space heritage described in section \ref{subsec:space_asic} use ASICs which are designed to operate alongside the detector they control at cryogenic temperatures. All other systems studied in section \ref{sec:previous_work} go to great length to ensure that detector control electronics operate near room temperature. 

Some further investigation into component classes most affected by low temperature operation is performed.

\subsubsection{Ceramic Capacitors}
Modern electronics rely on the use of class II dielectric materials like X5R, X7R {(letters/numbers encoding operating temperature range and initial tolerance)} in capacitors. Those materials allow integration of several microfarads of capacitance in a 1mm by 0.5mm package. With high-speed digital electronics the use of such devices is considered mandatory as noise bypass capacitors. Typically rated for operation down to $-55^\circ C$ by the manufacturer, a study\cite{Teyssandier2010} shows that at 77K both capacitance and equivalent series resistance are degraded. Silicon capacitors are available\cite{Darcy2025} that solve low temperature operation issues. Widely available models do not yet match the energy density or cost-effectiveness of class II ceramic capacitors at the time of publication, but the technology is worth observing.

\subsubsection{Inductors}
The use of inductors is required for the implementation of switch mode power converters to generate the local supply voltages for the electronics as well as for isolation transformers to galvanically isolate the detector controller preventing the injection of {direct current (DC)} into the detector. Yin et al. \cite{Yin2021} show that modern nanocrystalline and amorphous materials work well at cryogenic temperatures. Previous work of the authors \cite{Richerzhagen2023} shows that it is possible to use those materials to construct practical inductive components but that, due to the manual winding techniques applied, the resulting components are very large and not suitable for small scale integration.

Commercially available inductive components typically limit their operational range to $-40^\circ C$. While it is possible to re-characterize electrical parameters with reasonable effort, without insights into proprietary manufacturing processes it is not possible to guarantee mechanical integrity of the part under thermal contraction.

\subsubsection{Integrated Circuits}
Both the automotive and industrial catalogues of major semiconductor manufacturers define the lower operating temperature limit at $-40^\circ C$. There is a much smaller selection of military grade components rated at $-55^\circ C$, but it may be missing required components, like DDR-SDRAM {(double data rate synchronous dynamic random access memory)} or Ethernet PHY {(physical layer controller)} capable of operation at $>1 Gbps$. 

For common electronics design methods, calculation and simulation, to work, all parameters need to be known and stable. If a component is to be used outside its specified range it needs to be fully retested, which is feasible for space grade electronics with the support of the manufacturer but not by external parties like ESO.

%%%%%%%   Networking   %%%%%%%
\subsection{Networking}
\label{subsec:dch_if_protocol}
As shown in section \ref{sec:previous_work} all large detector controller systems studied use a stream or packet based high-speed link to communicate between system components close to the focal plane and those further away. Standard gigabit Ethernet is widely used as shown in table \ref{tab:overview}. Another solution is the use of a custom protocol or camera link via optical fiber.

Within the ESO internal engineering standards there is a clear preference for Ethernet based standards. With further advantages being free choice between fiber based and twisted pair copper cable physical layer and available standards to embed synchronization and power (for copper based links) into the same link, only Ethernet based links are further studied for use in future ESO systems. 

\subsubsection{Network Requirements}
As a first order estimate for required bandwidth for image data transmission in a science application it is assumed that a typical $4K \times 4K$ detector is read-out at 1 fps continuously with 16-bit pixel depth. This yields $4096px \cdot 4096px \cdot 1 fps \cdot 16 bit = 268 Mbps$ net bandwidth.

Absolute data integrity (zero packet loss) is mandatory for science data and control commands. Traffic management has to handle large image transfers bursts following long idle periods and congestion control with the ability to assert back-pressure is needed when network saturation cannot be avoided. During detector readout extreme data bursts can overwhelm standard switch buffers, leading to packet loss, unless mitigated by effective quality of service (QoS) strategies. Additionally, the system must support coordinated traffic shaping across multiple detectors to prevent simultaneous bursts from saturating network links. 

\subsubsection{Protocols}
The transmission control protocol (TCP) and higher level protocols based on it provide a reliable communication channel guaranteeing receipt of all packets in the correct order, but requires a complex state machine and buffer memory that may be hard to implement on low level hardware lacking a full central processing unit (CPU). Nonetheless, Hyper Suprime-Cam successfully uses TCP based protocols to communicate\cite{Miyazaki2018} implemented by a compact FPGA core, SiTCP\cite{Uchida2008}. If further system design shows that the integration of a TCP core is not feasible in the specific application, lower level protocols can be considered.

The real time MUDPI streaming protocol (RTMS)\cite{Suarez2023} used on NGCII is based on the { user datagram protocol (UDP)} and a strong option for future ESO detector control systems. The packets are fully routable and can be transmitted over a standard real-time capable network infrastructure all the way to the final consumer. No QoS considerations are made by the protocol and QoS needs to be guaranteed by the network. Standard enterprise switches provide well-established QoS mechanisms that can address many of the above-mentioned requirements. Features like differentiated services codepoint (DSCP) marking (RFC2474\cite{ietf_rfc2474}), priority queuing (IEEE 802.1Q\cite{ieee_802.1Q_2022}), and weighted fair queuing offer reliable traffic classification and basic congestion management. However, several limitations become apparent with increased array count. The relatively small buffers of enterprise switches may be insufficient for simultaneous detector bursts. Without effective back-pressure, uplinks can saturate and queues will overflow. Standard traffic management can cause link starvation where some detectors monopolize bandwidth while others drop packets. Coordinating trigger timing across hundreds of endpoints requires application-layer control beyond basic switch features.

GigE-Vision\cite{gige-vision} is an industrial imaging protocol stack consisting of three UDP based protocols. GigE-Vision has an optional frame-retransmit feature which would require some buffer memory on the detector control hardware. GigE-Vision can be routed all the way to the final consumer. GigE-Vision provides good compatibility with standard QoS implementations due to its structured approach separating control (GVCP{, GigE Vision Control Protocol}) and streaming (GVSP{, GigE Vision Stream Protocol}) traffic. This separation enables straightforward priority marking and queuing. The protocol also includes rate-controlling features like stream channel bandwidth limiting, frame rate throttling and packet size control. However, some aspects present challenges. The optional packet resend mechanism requires substantial buffering that may exceed simple detector control hardware capabilities, UDP-based streaming can overrun buffers during coordinated detector readouts, and global bandwidth arbitration across hundreds of detectors requires aggregator-level coordination beyond the protocol's original scope. 

Another option to be explored is the use of layer 2 protocols. Layer 2 protocols are not routable and limited to the local link. While the embedded firmware can be very simple, a data aggregator is needed to translate the layer 2 protocol to a routable protocol. In that, layer 2 protocols are similar to the custom data links used by many instruments (see table \ref{tab:overview}) but based on the Ethernet physical layer. If some edge compute features are integrated with the aggregator{,} it would fulfil a similar role to the SDS of the LSST Camera.

\subsubsection{Power Over Ethernet}
Power over Ethernet {(PoE)} is a common feature of office networks and widely supported by network equipment. It allows the injection of power into copper-based Ethernet links. Several profiles are available, with two basic modes listed in table \ref{tab:poe}. Higher power profiles are available but not considered necessary for highly integrated detector controller applications.

\begin{table}[ht]
\caption{Power over Ethernet Profiles} 
\label{tab:poe}
\small
\begin{center}
\begin{tabular}{|l|l|l|l|l|}
\hline
\rule[-1ex]{0pt}{3.5ex} IEEE Standard & Common Name & Pairs & Power at Sink & Voltage at Sink\\ \hline \hline
\rule[-1ex]{0pt}{3.5ex} 802.3af\cite{ieee_802.3_2022} & PoE  & 2 & 12.95W & 37V to 57V\\ \hline
\rule[-1ex]{0pt}{3.5ex} 802.3at\cite{ieee_802.3_2022} & PoE+ & 2 & 25.50W & 42.5V to 57V\\ \hline
\end{tabular}
\end{center}
\end{table} 

Power over Ethernet controller ICs are widely available including example designs that can be followed. Ethernet uses isolation transformers for the communication lines and the main power converter, converting the $+48V$ nominal PoE voltage to usable lower voltages {and} is generally isolated as well. This means that, even with copper based Ethernet, galvanic isolation of the detector electronics can be maintained which is required for the implementation of effective detector protection from DC current injection and low frequency current loops.

\subsubsection{Physical Layer Options}
For Ethernet data communication over four-pair twisted pair cables the standards listed in table \ref{tab:ethernet} are available. Legacy standards too slow for the application are not listed.

\begin{table}[ht]
\caption{List of Twisted Pair Ethernet Standards} 
\label{tab:ethernet}
\small
\begin{center}
\begin{tabular}{|l|l|l|l|l|l|l|l|l|l|}
\hline
\rule[-1ex]{0pt}{3.5ex} IEEE Standard & Common Name &Speed [Mbps] &  Cable \\ \hline \hline
\rule[-1ex]{0pt}{3.5ex} 802.3ab\cite{ieee_802.3_2022}& 1000BASE-T & 1000  & CAT5 \\ \hline
\rule[-1ex]{0pt}{3.5ex} 802.3bz\cite{ieee_802.3_2022}& 2.5GBASE-T & 2500  & CAT5e\\ \hline
\rule[-1ex]{0pt}{3.5ex} 802.3bz\cite{ieee_802.3_2022}& 5GBASE-T   & 5000 & CAT6 \\ \hline
\rule[-1ex]{0pt}{3.5ex} 802.3an\cite{ieee_802.3_2022}& 10GBASE-T  & 10000 & CAT6A \\ \hline
\end{tabular}
\end{center}
\end{table} 

{With Ethernet PHY the media independent interface (MII) and media dependent interface (MDI) are distinguished.} 5GBASE-T and 10GBASE-T require a complex {MII} between physical layer controller (PHY) and medium access layer controller (MAC), consisting of four high-speed lanes. On top of functional complexity, this will increase power consumption of PHY and FPGA and due to the limited number of suppliers availability of a low temperature rated part becomes less likely.

From the remaining modes, 1000BASE-T is ubiquitous and supported by nearly all modern network equipment. PHY are widely available. 2.5GBASE-T is also widely used nowadays, and it can operate with an overclocked single-lane MII interface compatible with 1000BASE-T. Compatible PHY are still much less available than for 1000BASE-T.

{While the twisted-pair based Ethernet standards listed in table \ref{tab:ethernet} are all intended for a maximum cable length of 100m, in practice the copper based network should be kept local. For the ELT this would mean limiting the local link to one instrument platform. This is to mitigate the effects of building-level disturbances like indirect lightning strike. Relevant product standards\cite{iec61326_1} reflect this common practice by not demanding surge voltage testing, simulating indirect lightning strike, on I/O lines installed indoors with a length of 30m or less. Preliminary test results on NGCII electromagnetic compatibility have not revealed any issues with immunity of twisted-pair Ethernet ports, so this standard is considered suitable for detector control applications.}

\subsubsection{Synchronization}
The ELT uses the precision time protocol (PTP) (IEEE 1588-2019\cite{ieee_1588_2019}) for time-stamping events on the telescope. PTP gives microsecond accuracy timestamps, but performance may be degraded on a congested link which is to be further evaluated. If there is the requirement to syntonize clock frequencies on several units, synchronous Ethernet (SyncE) (ITU-T G.8262\cite{itu-t_g.8262}) can derive a clock from the recovered Ethernet data clock. Combining both approaches and adding link calibration yields White Rabbit\cite{Lipinski2011} which can give sub-nanosecond absolute timestamp accuracy on a dedicated link but may degrade on a shared network link.

%%%%%%%  Edge Computing %%%%%%%
\subsection{Edge Computing}
\label{subsec:edge_computing}
The standard approach at ESO for image data processing beyond basic pixel sorting and accumulation is to use a detector workstation implemented with standard server PC hardware. For reading hundreds of detectors it needs to be investigated if this approach is still feasible when processing \textgreater100Gbps of image data. 

The LSST Camera {chose} to use an ATCA\cite{Atca} based science data system for image processing instead\cite{Kahn2010}. Space based systems need to reduce and compress image data before transmission to earth.

In general the concept of placing compute resources closer to the entities generating data is called ``edge computing'', at the edge of a network\cite{Cao2020}. For future detector controller systems it should be investigated if moving compute resources from the data center to the edge of the network is advantageous.

%%%%%%%%%%%%%%%%%%%%%%%%%%%%%%%%%%%%%%%%%%%%%%%%%%%%%%%%%%%%%
%%%%%%%                Proposed Concept              %%%%%%%%
%%%%%%%%%%%%%%%%%%%%%%%%%%%%%%%%%%%%%%%%%%%%%%%%%%%%%%%%%%%%%
\section{Proposed Concept of the HYDRA Controller}
\label{sect:concept}
Learning from the concepts studied in section \ref{sec:previous_work} and implementing technologies discussed in section \ref{sec:technologies} we propose a HYpothetical Detector Readout Array (HYDRA). We invite instrument consortia in early design phases to discuss the concept with us and consider it for large detector systems where NGCII is not suitable.

As top level requirement we aim to control a fully-digital detector or analog detector with existing companion ASICs. A simple use case of adding a low number of ADCs is also considered. We assume that high level system engineering can integrate the detector control hardware close to the detector using rigid-flexible PCB assemblies and that space can be reserved inside the vacuum vessel at $-40^\circ C$ or above. We plan to limit the power dissipation of the detector system, including the detector itself, up to the first packet based long-range data link to 15W, per detector. To limit the amount of external cabling required we set the initial requirement to use a single four-pair twisted pair cable for power, data and synchronization.

\begin{figure}[ht]
\centering
\includegraphics[width=\textwidth]{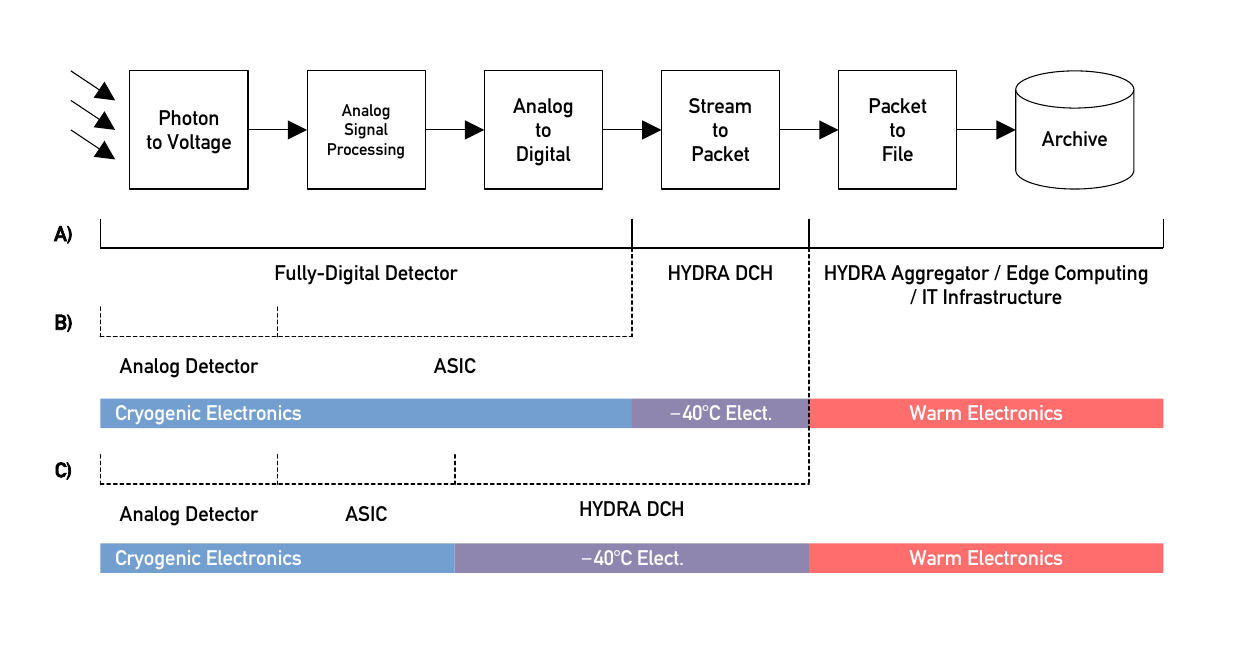}
\caption{Detector Control Model, HYDRA}
\label{pic:det_model_hydra}
\end{figure}

In our detector control model we place HYDRA as shown in figure \ref{pic:det_model_hydra}. If a fully digital detector meeting instrument requirements is available we propose its use (option A). Alternatively an analog detector can be adapted to a fully digital interface using an existing companion ASIC (option B). If this is not an option, for example if analog processing ASICs are available but not integrating ADCs, commercial off-the-shelf ADCs can be placed on the HYDRA { Detector Control Hardware} (DCH) (option C). This should be limited to a low number of ADCs (e.g. four) to maintain concept feasibility.

\begin{figure}[ht]
\centering
\includegraphics[width=\textwidth]{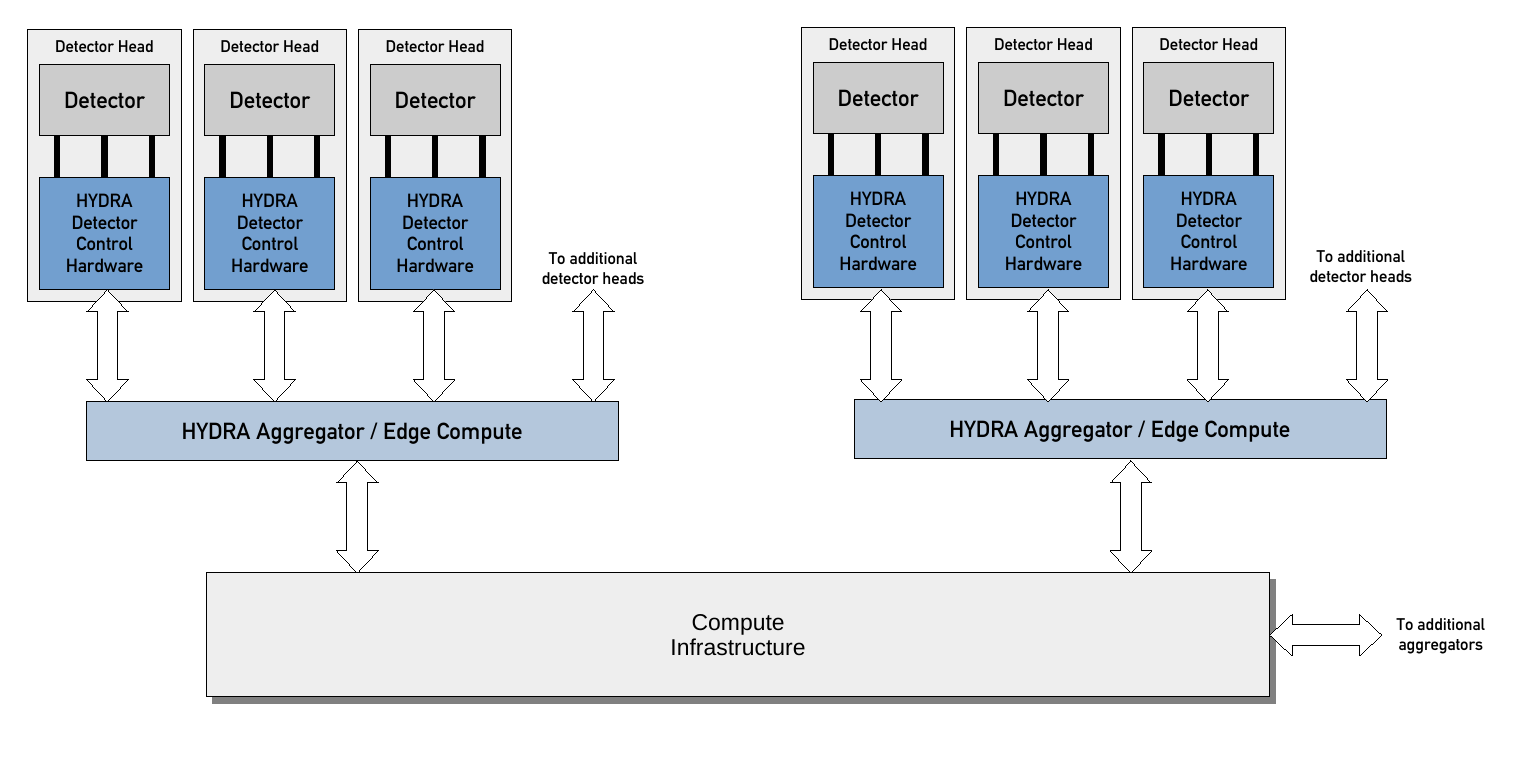}
\caption{Proposed System Block Diagram}
\label{pic:sys_bd}
\end{figure}

All options listed above contain common functionality, called HYDRA core. We propose to bundle all outside interfaces into a single, Ethernet based link, aggregating communication, synchronization and power transmission. This minimizes cabling to a four pair Ethernet cable for which vacuum compatible cables and connectors are available. Outside the vacuum vessel a HYDRA aggregator is placed that injects power and synchronization signals into the aggregated link and combines data streams from multiple HYDRA detector control hardware units into a single uplink to {the} central compute {system}. We anticipate that one aggregator could serve 8 to 32 detector control hardware units dependent on the outcome of de-risking activities described in section \ref{subsec:derisk}. The architecture of central compute infrastructure is highly dependent on the instrument design and cannot be discussed here. The aggregator can include edge-computing features if it is found to be beneficial to the overall system architecture. A block diagram is shown in figure \ref{pic:sys_bd}.

\subsection{HYDRA Detector Control Hardware}
\label{subsec:hydra_dch}
\begin{figure}[ht]
\centering
\includegraphics[width=0.7\textwidth]{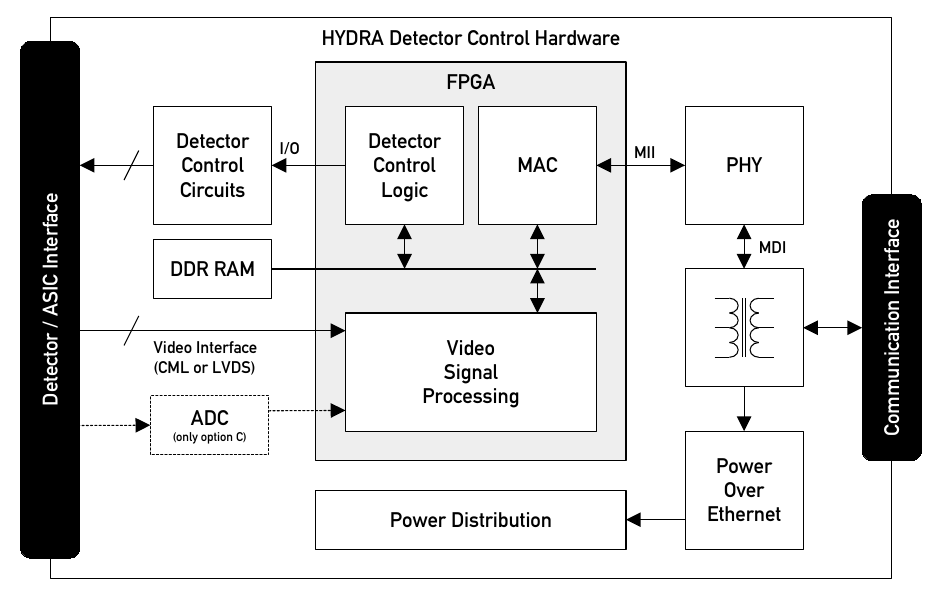}
\caption{Proposed Detector Control Hardware Block Diagram}
\label{pic:dch_bd}
\end{figure}

The proposed detector control hardware for the HYDRA system is described in more detail in this section. A block diagram of the detector control hardware is shown in figure \ref{pic:dch_bd}.

A relatively simple FPGA, for example from the Xilinx Spartan Ultrascale+ or Artix Ultrascale+ family in a 19x19mm package is considered sufficient. DDR memory is required for buffering at least two detector images. Double buffering allows reading into one buffer and transmitting from the other. When a new detector image is fully read, the read and transmit buffer are swapped. For a $9K\times9K$ detector the required memory space at 16bit pixel depth is calculated.

\begin{equation}
  2 \cdot 9000px \cdot 9000px \cdot 2Byte = 324MByte
\end{equation}

With the pixel depth at 16bit, it makes sense to implement a 16bit memory interface. With widely available DDR3L memory, 1Gbyte can be implemented with one or two memory chips. The available memory bandwidth at a conservative 400MHz memory clock rate, is:

\begin{equation}
  800MT/s \cdot 16bit = 12.8Gbps
\end{equation}

This is a factor x10 faster than the planned communication interface bandwidth and thus sufficient.

\subsubsection{Form Factor}
It is proposed to implement the entire detector control hardware on a single rigid-flexible PCB. This allows adaption to a complex mechanical envelope behind the detector or near it. A single PCB assembly allows full electrical testing after production using automated equipment. At this stage of the concept we do not propose a modular hardware design but enable design re-use by defining clean board-internal interface between functional blocks. Board-to-board connectors, especially for numerous high-speed differential signals, would inflate design volume. A continuous PCB allows the implementation of close impedance control and shielding as also highlighted in ref.\citenum{Onaka2008}.

\subsubsection{Power Budget}
A top level power budget for the detector control hardware is calculated in table \ref{tab:power_budget}. Power estimates are drawn from literature, where available, or estimated from NGCII and other experience at ESO.

\begin{table}[ht]
\caption{Power budget} 
\label{tab:power_budget}
\footnotesize
\begin{center}
\begin{tabular}{|l|c|c|c|l|}
\hline
\rule[-1ex]{0pt}{3.5ex} Component                               & Option A & Option B  & Option C & Source\\ \hline \hline
\rule[-1ex]{0pt}{3.5ex} \multirow{2}{*}{Detector}               & 2.6W  & -     & -    & Average from table \ref{tab:detectors} \\ \cline{2-5}
\rule[-1ex]{0pt}{3.5ex}                                         & -     & 0.1W  & -    & Estimate for HAWAII-4RG\\ \hline
\rule[-1ex]{0pt}{3.5ex} \multirow{2}{*}{Analog Signal Proc.}    & -     & 0.3W  &  -    & Design requirement for ACADIA\cite{Loose2018} \\ \cline{2-5}
\rule[-1ex]{0pt}{3.5ex}                                         & -     & -       & 2.7W & 1/9x LSST FEC\cite{OConnor2012}\\ \hline
\rule[-1ex]{0pt}{3.5ex} ADC                                     & -     & -     & 1.0W & Estimate for 4x LTC2387\\ \hline
\rule[-1ex]{0pt}{3.5ex} PHY                                     & \multicolumn{3}{|c|}{0.5W} & Estimate for 1GbE PHY \\ \hline
\rule[-1ex]{0pt}{3.5ex} DDR Memory                              & \multicolumn{3}{|c|}{0.3W} & Estimate for 1GByte DDR3L\\ \hline
\rule[-1ex]{0pt}{3.5ex} FPGA                                    & \multicolumn{3}{|c|}{4.0W} & Estimate for Xilinx XCAU10P\\ \hline
\rule[-1ex]{0pt}{3.5ex} Misc.                                   & \multicolumn{3}{|c|}{0.5W} & Estimate\\ \hline 
\rule[-1ex]{0pt}{3.5ex} Loss                                    & 2.4W  & 1.7W  & 2.7W & $\eta = 0.7$\\ \hline \hline
\rule[-1ex]{0pt}{3.5ex} Power at plug                           & 10.3W  & 7.4W  & 11.6W & \\ \hline
\end{tabular}
\end{center}
\end{table} 

The configurations differ marginally. In any case the power consumed is suitable for power delivery through PoE using one of the basic profiles listed in table \ref{tab:poe}. For use in instrument power budget, beyond a simple sustainability evaluation and very basic conceptual evaluation of cooling systems, more detailed calculations for the specific application are required.

\subsection{Hydra Aggregator}
The functionality of the block marked HYDRA aggregator in figure \ref{pic:sys_bd} is highly dependent on the outcome of further de-risking activities. A minimum viable configuration could be a commercial-off the shelf network switch that is able to inject PoE power and a PTP boundary clock. This assumes that the QoS measures implemented on the switch and the protocol chosen for detector control hardware communication can guarantee the necessary low error rate.

If more complex protocol translation or edge computing features are needed, a more complex aggregator like the SDS of the LSST camera are chosen. It could be based on a modular standard like ATCA\cite{Atca} or MicroTCA\cite{Microtca} or a ground-up custom development.

{
\subsection{Cost}
Per-unit hardware cost for the HYDRA core DCH, as described in section \ref{subsec:hydra_dch}, is estimated to be lower than €5000, derived from experience with NGCII production. This is significantly lower than cost of a modular detector controller and insignificant compared to the cost of a detector.

For the aggregator and compute infrastructure the assumed baseline is a basic, commercial Ethernet switch and server-based central compute facility as for current instruments. From that baseline, functionality can be re-balanced in a cost conscious way, for example moving compute resources to the edge so they can be omitted in the data center. 

The key element regarding cost is that of the detector and commercial or custom companion ASIC, if required. The HYDRA concept can help reduce compounding cost effects of cabling, cabinets, size of instrument structure, power infrastructure compared to other systems like NGCII.}

\subsection{De-Risking Activities}
\label{subsec:derisk}
Once interest is expressed from the community, some de-risking activities can be started. Those are required before committing to formal requirements. The goal of these activities is to increase the technology readiness level {(TRL)} of technologies discussed in section \ref{sec:technologies}. 

{ESA\cite{ESA_TRL} (European Space Agency) and NASA\cite{NASA_TRL} (National Aeronautics and Space Administration) use in-house definitions for TRL ranging from ``basic principle'' (TRL-1) to ``flight proven'' (TRL-9) with minor deviations between the two agencies. Due to the space focus of both agencies, the TRL definitions are not well suited to assess technology progression in ground based astronomy. ESO internal TRL definitions could be used, but for the purpose of this paper we use the more general TRL definitions from the European Union's Horizon Europe program\cite{trl2025}, also reproduced in table \ref{tab:trl_definitions}.}

\begin{table}[ht]
\caption{TRL definitions, reprinted from Ref.\citenum{trl2025}} 
\label{tab:trl_definitions}
\footnotesize
\begin{center}
\begin{tabular}{|l|l|}
\hline
Level & Definition \\ \hline \hline
TRL-1 & Basic principles observed \\ \hline
TRL-2 & Technology concept formulated \\ \hline
TRL-3 & Experimental proof of concept \\ \hline
TRL-4 & Technology validated in a lab \\ \hline
TRL-5 & Technology validated in a relevant environment \\ \hline
TRL-6 & Technology demonstrated in a relevant environment \\ \hline
TRL-7 & System prototype demonstration in an operational environment \\ \hline
TRL-8 & System completed and qualified \\ \hline
TRL-9 & Actual system proven in an operational environment \\ \hline
\end{tabular}
\end{center}
\end{table}

While all technologies in section \ref{sec:technologies} are mature and proven in an operational environment (TRL-9) on the global scale, evaluation is done in context of the specific application environment at ESO with estimated TRLs before and after de-risking in table \ref{tab:trl_improvements}. TRL in the ``before'' column therefore refers to estimated TRL at ESO within the environment of HYDRA before de-risking. 

We aim to select de-risking activities in such way that TRL-6 (technology demonstrated in a relevant environment) is achieved before committing to a full development. {The ``delta'' column indicates the improvement in TRL required during de-risking.} We assume that the relevant environment for HYDRA encompasses 8 to 32 detector control hardware units operating in vacuum and at final temperature on a single aggregator with synthetic detector data representative of the readout of real detectors.

\begin{table}[ht]
\caption{Technology readiness level improvement} 
\label{tab:trl_improvements}
\footnotesize
\begin{center}
\begin{tabular}{|l|l|l|l|l|}
\hline
\rule[-1ex]{0pt}{3.5ex} Technology                          & Before & After & Delta & Comment \\ \hline \hline
\rule[-1ex]{0pt}{3.5ex} Fully Digital Detectors             & TRL-6  & TRL-6 & -  & \multirow{2}{*}{No activities planned}\\ \cline{1-4}
\rule[-1ex]{0pt}{3.5ex} ASICs                               & TRL-3  & TRL-3 & -  & \\ \hline
\rule[-1ex]{0pt}{3.5ex} Cold Electronics                    & TRL-5  & TRL-6 & +1 & \\ \hline
\rule[-1ex]{0pt}{3.5ex} Power over Ethernet                 & TRL-4  & TRL-6 & +2 & \\ \hline
\rule[-1ex]{0pt}{3.5ex} \multirow{3}{*}{Network Protocols}  & TRL-4 (for TCP) & \multirow{3}{*}{TRL-6} & +2 & \\ \cline{2-2} \cline{4-4}
\rule[-1ex]{0pt}{3.5ex}                                     & TRL-5 (for RTMS) &  & +1 & \\ \cline{2-2} \cline{4-4}
\rule[-1ex]{0pt}{3.5ex}                                     & TRL-4 (for GigE Vision) & & +2 & \\ \hline
\rule[-1ex]{0pt}{3.5ex} Synchronization                     & TRL-4  & TRL-6 & +2 & \\ \hline
\rule[-1ex]{0pt}{3.5ex} Edge Computing                      & TRL-2  & TRL-4 & +2 & \\ \hline
\end{tabular}
\end{center}
\end{table} 

Running a fully digital detector has been demonstrated with NGCII using the {GeoSnap-18} ROIC\cite{Richerzhagen2025}. The SIDECAR ASIC has been integrated with ESO detector controllers in an experimental setup\cite{Dorn2010}. An initial demonstration of a custom {preamplifier} ASIC for analog infrared CMOS detector was done in 2018\cite{Ackermann2018}. No risk reduction activities for those technologies are planned in the scope of the detector controller but should be conducted in parallel, with electro-optical properties of the detector in mind.

Cold electronics development for operation at $-40^\circ C$ or $-55^\circ C$ can be brought to TRL-6 by pre-development of detector control hardware core functionality. Cold electronics are a known technology at ESO, but integration of more complex digital components needs to be proven. This should include making a selection of HYDRA DCH core components, FPGA, PHY, PoE power converter and DDR memory and construction of a prototype DCH unit on a realistic PCB stack-up but disregarding form-factor. Extensive thermal testing of multiple units at low temperature should be performed. Additionally, highly accelerated lifetime tests can be performed to identify weak points early. This pre-development would also bring PoE technology, which is not used in ESO detector systems so far, to TRL-6. Another outcome of this activity is a decision if operation at $-55^\circ C$ is possible or if the temperature range needs to be limited to $-40^\circ C$.

Using this prototype hardware, the feasibility of implementing TCP based protocols on the FPGA can be evaluated, and a protocol stack can be chosen including the initial testing of switches. The simultaneous arrival of TCP or UDP image data packets on all ports of a switch can be tested and evaluated for packet loss. This can determine if commercial off-the-shelf switches are suitable or if a custom development for the aggregator is required. Network based synchronization methods like PTP or a full or partial implementation of White Rabbit can be demonstrated under realistic conditions as well.

It is also required to study if edge computing is beneficial for the overall detector system, what are the required changes to ESO standard software pipelines, and if some pipeline steps are better run on an FPGA or {graphics processing unit (GPU)} rather than a CPU. The evaluation of new software concepts is considered an activity of similar magnitude as hardware and firmware development. The proposed concept for the system allows full flexibility regarding compute resources, so this development can be done in parallel. It is unlikely that edge computing technology can be evaluated all the way to TRL-6 in the same time-frame as other de-risking activities but TRL-4 (technology evaluated in a lab) should be achievable.

\subsection{Full Development}
If a full development is started, first the ESO systems engineering processes need to be applied, starting with collection of formal requirements. The key requirement is the final selection of the detector(s) to be supported.

With the outcome of the de-risking activities, conceptual design can be finalized, making a final decision on the scope of the aggregator and the interface between HYDRA DCH and aggregator. Work-packages can be set up for system engineering, DCH hardware development, aggregator procurement or development as well as firmware and software.

{

\subsection{Summary and Outlook}
To summarize the proposal of HYDRA as a detector controller for large ESO instruments, in this section we will outline specific points in the design we intend to emphasize during the development, point out possible tradeoffs we might consider and weigh the importance of design decisions made in this section.

Regarding detector support we try to keep the choice as open as possible for instrument designers to not impede high-level trade-offs. While there are no concerns regarding fully-digital detectors and detectors with an existing companion ASIC handling digitization, we consider the integration of CCDs to be more problematic. To support CCDs, specifically those with more than four video outputs, we anticipate that development of an ASIC needs to be considered.

Simplifying the main interface of the detector control hardware is a key premise of the HYDRA concept. Should de-risking activities show this is entirely unfeasible, the HYDRA concept as a whole needs to be questioned. A possible tradeoff would be allowing for two Ethernet interfaces per DCH, for example to separate timing sensitive traffic from bulk image data. The choice of protocol is completely open, and we cannot anticipate the outcome of the proposed tests.

For operating temperature of the DCH we consider $-40^\circ C$ a hard lower limit at the moment with a slight chance that operation at $-55^\circ C$ might be possible with extra effort. However, many designs discussed in section \ref{sec:previous_work} integrate warm electronics inside the cryostat, accepting the thermal gradient between detector and controller. From this we speculate that final operating temperature of HYDRA may be at a level higher than $-40^\circ C$, possibly even without any active temperature control as on Hyper Suprime-Cam.

The final implementation of the aggregator and compute infrastructure is highly dependent on the outcome of internal and external activities on system and software level, so no estimation can be given at this time.
}

\section{Conclusion}
\label{sect:conclusion}
We conclude that in ground-based astronomy there is a wide variety of solutions to the problem of detector control at a large scale with some common concepts observed. There is a preference to run detector control electronics within the temperature range covered by the commercial and industrial catalogues of component manufacturers. Electronics that need to run at lower temperatures are generally implemented by custom ASICs. Furthermore, it is observed that Ethernet based communication links, specifically 1GbE, are the most common solution for linking system components close to the detector with those further afield.

Concepts discussed in section \ref{sec:previous_work}, specifically the LSST Camera, Pan-STARRS and Hyper Suprime-Cam, show that detector control electronics are integrated into the system and were clearly considered from an early point in conceptual design of the instrument. We consider this approach mandatory for the success of large detector systems and hope to contribute to an early discussion regarding future large ESO instruments with this paper.

We further describe key technologies for large detector systems, namely fully-digital detectors as an alternative to custom ASICs and Ethernet based communication links, perform a short analysis of temperature limits for electronics and conclude that $-40^\circ C$ is a safe lower design limit with a possibility to extend that range to $-55^\circ C$ when limiting component choice to military grade components.

Finally, we propose a  concept for a hypothetical detector controller that can be integrated into a detector system that is ideally suited to control a fully-digital detector but can also be expanded to control analog detectors with a companion ASIC or even analog CCDs. Our concept relies on a single Ethernet link aggregating communication, synchronization and power features.

\subsection* {Code, Data, and Materials Availability} 
The data that support the findings of this study are available from the corresponding author upon reasonable request.

\subsection* {Disclosures} 
The authors declare there are no financial interests, commercial affiliations, or other potential conflicts of interest that have influenced the objectivity of this research or the writing of this paper.

\bibliography{report}
\bibliographystyle{spiejour}

%%%%% Biographies of authors %%%%%

\vspace{2ex}\noindent    \textbf{Mathias Richerzhagen} is a detector electronics engineer at the European Southern Observatory. He received his engineering diploma from RWTH Aachen University in 2012. His current research includes development of the detector controller for the ELT as well as some work in cryogenic electronics.

\vspace{2ex}\noindent    \textbf{Naidu Bezawada} is a detector engineer at ESO, currently leading the detector systems work package for MICADO. Working in close collaboration with the astronomical community and with the detector manufacturers, he has contributed to many astronomical detector systems for more than 30 years, at the Indian Institute of Astrophysics, Bengaluru, India, at the UKATC in Edinburgh and ESO in Germany.

\vspace{2ex}\noindent    \textbf{Sebastian Egner} is a systems engineer at the European Southern Observatory. He has worked on various optical and near-infrared instruments for Earth observation and ground-based astronomy, including Adaptive Optics systems over the past 20 years.

\vspace{2ex}\noindent   \textbf{Elizabeth George} has worked on a variety of instruments and detector systems for astronomy from the millimeter to visible wavelengths throughout her career. She is currently a detector engineer at ESO working on visible and infrared detector systems for instruments on ESO’s Very and Extremely Large Telescopes and future facilities.

\vspace{2ex}\noindent   \textbf{Alessandro Meoli} is a detector engineer at the European Southern Observatory, working on visible and infrared detector systems for ESO’s Extremely Large Telescope. He is currently leading the detectors work package of the MOSAIC instrument. He contributes to the calibration and validation of the ELT wavefront sensing cameras and is currently involved in the development of curved CMOS detectors and in investigations on cryogenic detector electronics. Prior to joining ESO, he worked at the European Space Agency on optoelectronic systems and instrument electronics for space missions, and at CERN on instrumentation development for high-energy physics experiments.

\vspace{2ex}\noindent   \textbf{Alexander Rüde} is a detector electronics engineer at the European Southern Observatory. He received his M.Eng. degree from the University of Applied Sciences Jena in 2018. Currently, his work focuses on firmware and FPGA development for the NGCII detector controller for  ESO’s future astronomical instruments.

\vspace{2ex}\noindent   \textbf{Matthias Seidel} is a detector electronics engineer at the European Southern Observatory. He received his engineering diploma from the University of Applied Sciences in Bocholt in 2010. Initially focusing on electronics and FPGA development in the field of XRF spectroscopy, he is currently working on ESO’s NGC and NGCII detector controllers

\vspace{1ex}
\noindent Biographies and photographs of the other authors are not available.

\listoffigures
\listoftables

\end{spacing}
\end{document}